\documentclass[twocolumn,           
               showpacs,            
               preprintnumbers,     
               aps,                 
               prd,          	    
               letterpaper,             
               groupaddress,      
               nofootinbib,         
               tightenlines,        
               floats,floatfix      
               ]{revtex4-1}

\usepackage{graphicx}
\usepackage{dcolumn}
\usepackage{bm}
\usepackage{amsmath}

\begin{document}
\preprint{APS/123-QED}

\title{Cosmological dark fluid from five-dimensional vacuum}

\author{Luz M. Reyes}%
 \email{luzreyes@fisica.ugto.mx}
\affiliation{%
Departamento de F\'isica, DCI, Campus Le\'on, Universidad de Guanajuato,
C.P. 37150, Le\'on, Guanajuato, M\'exico.}

\author{Jos\'e Edgar Madriz Aguilar}
 \email{jemadriz@fisica.ugto.mx}
\affiliation{%
Departamento de F\'isica, DCI, Campus Le\'on, Universidad de Guanajuato,
C.P. 37150, Le\'on, Guanajuato, M\'exico.}

\author{L. Arturo Ure\~na-L\'opez}%
 \email{lurena@fisica.ugto.mx}
\affiliation{%
Departamento de F\'isica, DCI, Campus Le\'on, Universidad de Guanajuato,
C.P. 37150, Le\'on, Guanajuato, M\'exico.}

\date{\today}

\begin{abstract}
In the framework of the induced matter theory of gravity, we derive a
5D solution of the field equations that can describe a 4D cosmological
scenario where the dark fluid (dark matter plus dark energy) equation
of state has a geometrical origin. There is not a natural separation
of the dark sector into different components, and then the model may
provide a geometrical explanation for the existence of a dark
degeneracy in cosmological scenarios.
\end{abstract}

\maketitle

\section{\label{sec:level1}Introduction}
The idea of the existence of exotic components in the universe, namely
dark matter and dark energy, comes basically from three observational
evidences: supernovae of the type Ia, surveys of clusters of galaxies,
and observations of anisotropies in the Cosmic Microwave Background
Radiation (CMBR) \cite{Olive:2010mh}. Observations of high redshift of
type Ia supernovae suggest that the expansion of the universe is accelerating\cite{Perlmutter:1998np,*Perlmutter:1997zf,*Riess:1998cb,*Barris:2003dq,Sullivan:2010zz,Dilday:2010xe,*Riess:2009zz}. Surveys of clusters of
galaxies indicate that the energy density of matter is less than the
critical density of the Universe\cite{Pope:2004cc}, whereas
observations of the temperature anisotropies of the Cosmic Microwave
Background (CMB) show evidence that our Universe is spatially flat,
and then the total energy density parameter is very close to the
critical one\cite{Larson:2010gs}.

In obtaining these observational constraints, it has been
considered the model assumption that there are two separated dark
components: dark matter, responsible for the formation of cosmological
structure, and dark energy, responsible for the accelerated expansion of the
Universe. From this arises the so-called standard cosmological model,
also known as $\Lambda$CDM, where dark matter is a (cold) pressureless
fluid and the dark energy is described by a cosmological
constant\cite{Olive:2010mh,*Benson:2010de}.

However, the observations from which these two components have been
detected are gravitational in nature, and these kind of measurements
are unable to provide information about a unique decomposition of the
dark sector into these
components\cite{Kunz:2009yx,*Hu:1998tj,*Wasserman:2002gb,*Rubano:2001su,*Liddle:2006qz}. As
it was clearly argued in\cite{Kunz:2007rk}, when we are in a state of
total ignorance about the nature of a single one of the dark
components, we can also not completely measure the others.

In this case, the separation into dark matter and dark energy can be
seen as a convenient parametrization without experimental reality. In
models where this split is assumed, it is necessary to impose some additional
conditions upon the models in order to make the assumption
well-defined. These conditions can be that the dark energy vanishes at
high redshift, or that the dark energy constitutes the non-clustering
part of the dark energy-momentum tensor. The lack of guidance of
gravity observations leads to the so-called dark
degeneracy\cite{Kunz:2007rk}.

Theoretically, a great effort has been done to construct models to
explain dark energy. In the literature, we can find proposals, among
others, like quintessence models, interacting models of dark energy,
k-essence models and proposals coming from modified theories of
gravity\cite{CalderaCabral:2008bx,*ArmendarizPicon:2000ah,*Capozziello:2005mj,*Durrer:2007re},
see also\cite{Copeland:2006wr,Tsujikawa:2010sc,Sapone:2010iz} and
references there in. Other alternatives are models based in theories
with extra dimensions, like the induced matter (IM) theory of
gravity\cite{Overduin:1998pn,Wesson:1999nq,*Wesson:2006ta} and Brane
World (BW)
scenarios\cite{Antoniadis:1998ig,*ArkaniHamed:1998rs,*Randall:1999ee,*Randall:1999vf,*Maartens:2010ar,*Yazdani:2011zz,*Brax:2003fv}. 

In BW scenarios, the 4D universe is viewed as a hypersurface called
the brane, which is embedded in a higher dimensional spacetime called
the bulk. In this context, ordinary matter is confined to the brane by
a variety of different mechanisms, while gravity can propagate freely
through the bulk. 

Parallel to BW scenarios we have the IM theory. This theory can be
considered as an extension of 4D general relativity to 5D. In this
approach, our Universe is described by a 4D hypersurface embedded in a
5D Ricci-flat ($^{(5)}R_{ab}=0$) spacetime. The extra dimension is
considered as non-compact, and classical sources of matter in 4D are
identified with the curvature of the 4D hypersurface. This curvature
is a consequence of the embedding\cite{Wesson:1999nq,*Wesson:2006ta}.

The IM theory is mathematically supported by the Campbell-Magaard
theorem, that states that any analytical solution of the
$n$-dimensional Einstein equations can be embedded in a
$(n+1)$-dimensional Ricci-flat
manifold\cite{magaard1:phd,*campbellbook:book,Lidsey:1997zx,Romero:1996ba}. Both the BW and
the IM theory have different physical motivations, an equivalence to
each other has been shown by Ponce de Leon
in\cite{PonceDeLeon:2001un}. In spite of such an equivalence, the
requirement in IM theory of starting from a 5D Ricci-flat spacetime,
makes the task of finding solutions easier than in the BW theories. 

The purpose of this paper is to derive a particular solution of the 5D
field equations of the IM theory capable to describe not just the dark
energy component of the universe, but the full dark sector (dark
matter $+$ dark energy) as a single dark fluid component. An exact
solution is found from the expected behavior of the dark fluid
equation of the state in the standard $\Lambda$CDM model. The solution
also unifies previous works in which dark matter and dark energy were
studied separately in IM theory.

The paper is organized as follows. In Sec.~\ref{sec:level2}, we obtain
the geometrical equation of state for the induced matter, from a class
of 5D solutions, and we establish the effective 4D field equations. In
Sec.~\ref{sec:level3}, we obtain a particular solution that describes,
in the sense of induced matter, the 4D cosmological dark
fluid. Finally, in Sec.~\ref{sec:level4}, we give some final
comments.

\section{\label{sec:level2} 5D solutions and 4D induced matter}

Let us start by considering a 5D Ricci-flat spacetime
($\mathcal{M}^5,g$). In order to describe neutral matter in the local
coordinates $\lbrace\xi^{a}\rbrace=\{ x^{\alpha},\psi\}$, we choose a
coordinate gauge that allows us to write the 5D line element in the
form
\begin{equation}
  \label{eq:a1}
  dS_{5}^{2} = g_{\alpha\beta}(x,\psi) \, dx^{\alpha}dx^{\beta} +
  \epsilon g_{\psi\psi}(x,\psi) \, d\psi^{2} \, ,
\end{equation}
where $\psi$ is the space-like and non-compact extra coordinate, and
parameter $\epsilon = \pm 1$ accounts for the signature of the
extra coordinate. Our conventions are: Latin (Greek) indices take
values $0,1,2,3,4$ ($0,1,2,3$), and the metric signature is
$(+,-,-,-,-)$, and we use units in which $c=1$. 

In IM theory the 4D field equations are in general constructed as
follows\cite{Wesson:1999nq,*Wesson:2006ta}. From the 5D field
equations $^{(5)}R_{ab} =0$, and by means of the Gauss-Codazzi-Ricci
equations for the embedding, we write the conventional 4D Ricci tensor
$^{(4)}R_{\alpha\beta}$. With the help of $^{(4)}R_{\alpha\beta}$ and
the induced metric $h_{\alpha\beta}(x) = g_{\alpha\beta}(x,\psi_0)$ on a
generic hypersurface $\Sigma: \psi = \psi_0 = \mathrm{constant}$, we
also write the 4D scalar of curvature, $^{(4)}R$. 

We can form the 4D Einstein tensor $^{(4)}G_{\alpha\beta} = \,
^{(4)}R_{\alpha\beta} -(1/2) \, ^{(4)}R \, h_{\alpha\beta}$, leaving
all remaining terms grouped together to form an effective (induced)
energy-momentum tensor $T^{(IM)}_{\alpha\beta}$. The resulting field
equations are then of the form $^{(4)}G_{\alpha\beta} = 8\pi G \,
T_{\alpha\beta}^{(IM)}$, where the energy-momentum tensor of induced
matter is purely geometrical in nature, and has the
explicit form\cite{Overduin:1998pn,Wesson:1999nq,*Wesson:2006ta}
\begin{widetext}
\begin{equation}
  8\pi G \, T_{\alpha\beta}^{(IM)} = \frac{\Phi_{,\alpha;\beta}}{\Phi}
  - \frac{\epsilon}{2\Phi^{2}} \left\lbrace
    \frac{\overset{\star}{\Phi}}{\Phi} \overset{\star}{g}_{\alpha\beta}
    - \overset{\star\star}{g}_{\alpha\beta}
   + g^{\lambda\mu} \overset{\star}{g}_{\alpha\lambda}
  \overset{\star}{g}_{\beta\mu} -\frac{1}{2}g^{\mu\nu}
  \overset{\star}{g}_{\mu\nu} \overset{\star}{g}_{\alpha\beta}
    + \frac{1}{4}g_{\alpha\beta}
    \left[\overset{\star}{g}^{\mu\nu} \overset{\star}{g}_{\mu\nu} +
      (g^{\mu\nu} \overset{\star}{g}_{\mu\nu})^{2} \right]
  \right\rbrace \, , \label{extra1}
\end{equation}
\end{widetext}
where $\Phi^{2}=g_{\psi\psi}$, the coma denotes partial derivative,
the semicolon denotes 4D covariant derivative, and the star $(\star)$
denotes partial derivative with respect to the extra coordinate
$\psi$.

In order to derive a cosmological scenario in which a combined dark
fluid may be described, let us use the class of 5D
solutions\cite{Liu:2001vm,Liu:1990pp,PonceDeLeon:1988rg}
\begin{equation}
  \label{eq:le}
  dS^{2}_{5} = [\dot{A}^2(t,\psi)/\mu^2(t)] dt^2 - A^2(t,\psi) \left( dr^2 +
    r^2d\Omega^2 \right) - d\psi^2 \, ,
\end{equation}
with
\begin{equation}
  A^2(t,\psi) = [\mu(t)^2 + k ] \psi^2 + 2 \nu(t) \psi +
  \frac{\nu(t)^2+K}{\mu(t)^2 + k} \, , \label{eq:pa}
\end{equation}
where the dot denotes derivative with respect to the time-like
coordinate $t$, $\mu(t)$ and $\nu(t)$ are arbitrary metric
functions, and $k$ is the 3D curvature constant. Constant $K$ is
related to the Kretschmann scalar, namely,
\begin{equation}
  \label{eq:kre}
  I = R_{abcd}R^{abcd} = \frac{72K^2}{A(t,\psi)^8} \, ,
\end{equation}
which shows that a non-null $K$ determines the Riemannian curvature of
the 5D manifold\cite{Zhang:2006qe}.

Assuming that the 5D spacetime can be foliated by a family of
hypersurfaces $\Sigma$, defined by the equation $\psi =
\mathrm{constant}$, the geometry of each hypersurface, say at
$\psi=\psi_0$, will be then determined by the induced metric
\begin{equation}
  \label{extra2}
  dS_{\Sigma}^{2} = [\dot{A}^2(t,\psi_0)/\mu^2(t)] dt^2 -
  A^{2}(t,\psi_0) \left( dr^2 + r^2 d\Omega^2 \right) \, .
\end{equation}  
The Friedmann-Robertson-Walker (FRW) metric for a 4D flat universe, as
supported by the observational evidence\cite{Larson:2010gs},
\begin{equation}
  \label{eq:le4d}
  dS_{|FRW}^2 = dt^2 - a^2(t) \left( dr^2 + r^2d\Omega^2 \right) \, ,
\end{equation}
where $a(t)$ is the scale factor of the Universe, can be recovered
from the induced metric~(\ref{extra2}) by demanding the continuity of
the metric across $\Sigma$. A comparison of Eqs.~(\ref{extra2}),
and~(\ref{eq:le4d}), indicate that the metric functions can be written
as
\begin{subequations}
  \label{eq:a}
  \begin{eqnarray}
    \mu(t) &=& aH \, , \label{eq:a3a} \\ 
    \nu(t) &=& -(aH)^2\psi_0 \pm \sqrt{-K+a^4H^2} \, , \label{eq:a4}
  \end{eqnarray}
\end{subequations}
where $H=\dot{a}/a$ is the cosmological Hubble parameter, and we have
set $k=0$. The double sign in Eq.~(\ref{eq:a4}) appears because the
Eqs.~(\ref{eq:pa}) and ~(\ref{eq:a3a}) leave to a quadratic equation
for the metric function $\nu$ when the continuity of the 5D metric is
imposed. Physically, both solutions for $\nu$ induce the same energy
density and pressure in 4D (see Eqs.~(\ref{eq:eff})), leaving to the
same 4D effective dynamics, and there is no dynamical reason to choose
any sign. 

As it is usually done in the IM theory, after the identification of
the induced matter on $\Sigma$ with a perfect fluid, we obtain, with
the help of Eqs.~(\ref{extra1}), (\ref{eq:le}), and~(\ref{eq:a}),
that the total effective energy density and isotropic pressure,
measured by 4D comoving observers, are given by\cite{Liu:2001vm}
\begin{subequations}
\label{eq:eff}
\begin{eqnarray}
  8\pi G \, \rho_{eff} &=& ^{(4)}G_0^0 = \frac{3\mu^2}{a^2} \,
  , \label{eq:dens} \\
  8\pi G \, p_{eff} &=& - ^{(4)}G_1^1 = -
  \frac{2\mu\dot{\mu}}{a\dot{a}} - \frac{\mu^2}{a^2} \,
  , \label{eq:pres}
\end{eqnarray}  
\end{subequations}
with $^{(4)}G_1^1 = ^{(4)}G_2^2 = ^{(4)}G_3^3$. By means of
Eqs.~(\ref{eq:a}), and~(\ref{eq:eff}), we can define the effective
equation of state
\begin{equation}
  \label{eq:a5}
  w_{eff} \equiv \frac{p_{eff}}{\rho_{eff}} = - \frac{1}{3} \left( 1 +
    2\frac{\dot{\mu}}{\mu H} \right) = - \frac{1}{3} \left( 1 +
    2\frac{\ddot{a}a}{\dot{a}^2} \right) \, ,
\end{equation}
which will give us information about the 4D physical sources of matter
that are induced geometrically on the hypersurface modeling our 4D
universe $\Sigma$. The last equality in Eq.~(\ref{eq:a5}) can be
recognized as the typical one between the effective equation of state
and the scale factor in a flat FRW Universe.

\section{\label{sec:level3} Inducing the Dark Fluid}
In the $\Lambda$CDM model, the evolution of the total dark sector
equation of state is given explicitly as
\begin{equation}
  w_{dark}(a) \equiv -\frac{\rho_\Lambda}{\rho_{\rm CDM} + \rho_\Lambda}
  = - \frac{a^3}{a^3+\alpha^3} \, , \label{eq:a6b} 
\end{equation}
where $\alpha = \sqrt[3]{\Omega_{\Lambda,0}/\Omega_{\rm CDM,0}}$ is
the ratio of the present values of the dark energy and dark matter
density parameters. In deriving Eq.~(\ref{eq:a6b}), we have considered
the convention for the normalization of the present value of the
scale factor, $a_0=1$, that we will use hereafter.

In order to give a geometrical interpretation in the context of IM
theory, of the dark fluid described by Eq.~(\ref{eq:a6b}), we
implement its formal identification with the geometrical equation of
state, see Eq.~(\ref{eq:a5}), so that
\begin{equation}
  \label{eq:a7}
  \frac{1}{3} \left( 1 + 2\frac{\ddot{a}a}{\dot{a}^2} \right) =
  \frac{a^3}{a^3 + \alpha^3} \, .
\end{equation}
A full integration gives the known functional form of the scale
factor $a(t)$ in a $\Lambda$CDM model,
\begin{equation}
  \label{eq:a16}
  a(t) = \alpha \sinh^{2/3}{ \left[ \frac{3}{2}
      \frac{H_0t}{\sqrt{1 + \alpha^3}} \right]} \, .
\end{equation}
This time, however, the matter contents only accounts for cold dark
matter and the cosmological constant. The substitution of the
result~(\ref{eq:a16}) into Eqs.~(\ref{eq:a}), allows us to find the
functional form of the metric functions,
\begin{subequations}
\label{eq:munu}
  \begin{eqnarray}     
    \mu(t) &=& \frac{\alpha^{5/2} H_0}{\sqrt{1 + \alpha^3}}
    \left(\frac{\cosh\gamma(t)}{\sinh^{1/3}\gamma(t)} \right) \,
    , \label{eq:mu} \\
    \nu(t) &=& - \frac{\alpha^{5} H_{0}^2}{1 + \alpha^3} \left(
      \frac{\coth^{2}\gamma(t)}{\sinh^{1/3}\gamma(t)} \right) \psi_{0}
    \nonumber \\ 
    && \pm \sqrt{-K + \left(\frac{\alpha^{7/3} H_{0}^2}{1 + \alpha^3}
      \right) \sinh^{2/3}\gamma(t) \cosh^{2}\gamma(t)} \, ,\label{eq:nu}
  \end{eqnarray}
\end{subequations}
where we have used the auxiliary function $\gamma(t) = (3/2)[H_{0}t /
\sqrt{1 + \alpha^3}]$.

Our results enclose some limiting cases already discussed in the
literature. On one hand, when $H_0 t \ll 1$, we find from
Eq.~(\ref{eq:mu}) that 
\begin{equation}
  \label{eq:a17}
  a(t) \propto t^{2/3} \, \Rightarrow \, \mu(t) \simeq \alpha \left(
    \frac{H_0}{\sqrt{1 + \alpha^3}} \right)^{2/3} \left( \frac{3t}{2}
  \right)^{-1/3} \, ,
\end{equation}
which corresponds to the cold 3D flat model studied within the IM
framework in Ref.~\cite{Liu:2001vm}. It can be seen that the effective
equation of state, see Eq.~(\ref{eq:a5}), corresponds to dust,
$p_{eff}=0$. Furthermore, we would like to point out that our solution
gives a full integration of the general case discussed
in\cite{Liu:2001vm} for a Universe containing dust and a cosmological
constant.

On the other hand, when $H_0 t \gg 1$, Eq.~(\ref{eq:mu}) yields
\begin{equation}
  a(t) \approx \alpha \exp \left( \frac{H_0t}{\sqrt{1 + \alpha^3}}
  \right) \, \Rightarrow \mu(t) \simeq \alpha H_0 \exp \left(
    \frac{H_0t}{\sqrt{1 + \alpha^3}} \right) \, . \label{eq:a18}
\end{equation}
This limiting solution corresponds to a 3D-flat universe dominated by
a cosmological constant, which has been extensively used in
Refs.~\cite{MadrizAguilar:2005tm,*MadrizAguilar:2006mv,PonceDeLeon:1988rg}
to derive inflationary scenarios from a non-compact Kaluza-Klein
theory of gravity.

\section{\label{sec:level4} Final Comments}
We have studied, within the context of an IM theory of gravity, a
cosmological scenario where the dark sector (dark matter plus dark
energy) can be geometrically induced upon our 4D universe from a 5D
Ricci-flat spacetime, as one single dark fluid. 

We were able to find a complete solution of the 5D metric functions
that covers both the matter and cosmological constant dominated epochs
in the evolution of the Universe. The result contains, as limiting
cases, solutions that have been derived and used by other authors also
in the context of IM theory. The central point of the geometrical
description relies on the formal identification of the geometrical
equation of state with the physical one, whose functional form is
given by the dynamics of the standard $\Lambda$CDM model. It should be
stressed out that only the dark matter and the dark energy components
were considered. Other matter components, as photons, baryons, and
neutrinos, are not part of the geometrical induced matter. However, 
these components can be included, if we extend our model to the context of hybrid models, in which the ideas of the IM theory and some ideas of brane cosmology can be combined. The idea of hybrid models has been introduced by J. Ponce de Leon in ~\cite{PonceDeLeon:2001un}. In these models the idea is to imposed the physics on the brane to restrict the freedom characteristic of many solutions in the IM theory. Hence, components like baryonic matter, neutrinos and photons could be introduced in the geometrical setting we are working on, in a similar way that it is done in brane cosmology, but this is matter of future investigation.

It can be seen from our results that the phenomenology associated to
a single dark fluid, for which there is no natural separation between
its components, can be given a geometrical interpretation from
the point of view of a non-compact Kaluza-Klein theory of
gravity. This fact may help to explain the existence of a dark
degeneracy which is supported by (gravitational) observations.

\begin{acknowledgments}
LMR and JEMA acknowledge support from CONACyT M\'exico. JEMA thanks
the kind hospitality of  the Departamento de F\'isica of the
Universidad de Guanajuato for a postdoctoral stay during which this
work was initiated. This work was partially supported by PROMEP, DAIP,
and by CONACyT M\'exico under grants 56946, and I0101/131/07 C-234/07
of the Instituto Avanzado de Cosmologia (IAC) collaboration.
\end{acknowledgments}

\bibliography{L4-refs}

\end{document}